\documentclass[dvips]{article}
\usepackage{icrctc07}

\title{Selection of Surviving Primary Protons at 4300 m a.s.l. with the ARGO-YBJ experiment}
\shorttitle{Surviving Primary Protons with ARGO-YBJ}

\authors{G. Di Sciascio$^{1}$ for the ARGO-YBJ Collaboration }
\shortauthors{G. Di Sciascio et al.}
\afiliations{$^1$ INFN, sez. di Napoli, Italy }
\email{giuseppe.disciascio@na.infn.it}

\abstract{The primary proton spectrum up to 100 TeV has been
investigated by balloon- and satellite-borne instruments. Above
this energy range only ground-based air shower arrays can measure
the cosmic ray spectrum with a technique moderately sensitive to
nuclear composition. An array which exploits the full coverage
approach at very high altitude can achieve an energy threshold
well below the TeV region, thus allowing, in principle, the
inter-calibration of the measured proton content in the primary
cosmic ray flux with the existing direct measurements from
balloons/satellites.

The capability of the ARGO-YBJ experiment, located at the
YangBaJing Cosmic Ray Laboratory (4300 m a.s.l., Tibet, P.R.
China), in selecting the surviving primary cosmic ray protons is
discussed. A procedure looking for quasi-unaccompanied events with
a very steep lateral distribution is also presented.}

\begin{document}
\maketitle

\section{Introduction}

Despite large progresses in operating new multi-component
Extensive Air Shower (EAS) experiments and in the analysis
techniques to infer energy spectra and chemical composition, the
key questions concerning the origin of the "knee" in the cosmic
ray energy spectrum are still open. In particular, one of the most
important questions to be solved is the position of the proton
knee (see for example \cite{pds} and reference therein). In fact,
different experiments have claimed to see a proton knee at
different energies: at 10 TeV the MUBEE collaboration
\cite{mubee}, at a few hundreds TeV the TIBET-AS$\gamma$
experiment \cite{tibet} and at a few PeV the KASCADE
\cite{kascade} and EAS-TOP \cite{eastop} experiments.
%
%
In addition, direct measurements carried out in the 100 TeV region
by RUNJOB \cite{runjob} and JACEE \cite{jacee}
do not exhibit any spectral break up to the highest measured
energy ($\sim$ 800 TeV).
The knowledge of the primary proton spectrum is of great
importance to understand the cosmic rays acceleration mechanisms
and propagation processes in the Galaxy. A careful measurement of
the proton spectrum over a wide range of primary energies (from
0.1 TeV to 10 PeV) using the same method is one of the main tasks
of the future cosmic ray experiments.

%
%
%
The energy region up to about 100 TeV has been investigated by
balloon- and satellite-borne instruments, and above these energies
only ground-based air shower experiments may provide data which
have however poor mass resolution. The detection of single hadrons
at ground level, strongly related to the surviving primary cosmic
ray protons, has been recognized for a long time as a method to
investigate the proton spectrum over a large energy range and to
derive the total inelastic cross section
\cite{yodh,eastop-psgl,kascade-psgl}.

The ARGO-YBJ experiment, an air shower array exploiting the full
coverage approach at the YangBaJing Cosmic Ray Laboratory (Tibet,
P.R. China, 4300 m a.s.l., 606 g/cm$^2$), offers the unique
opportunity to investigate the cosmic ray spectrum over a large
energy range (about 3 decades) because of its ability to operate
down to a few TeV, thus overlapping the direct measurements, by
measuring small size air showers (strip or digital read-out with
high spatial granularity) and up to the PeV region by measuring
the RPCs charge (analog read-out \cite{argo_bigpad}). In this
paper we will discuss how a direct measurement of the primary
proton spectrum could be achieved up to the PeV energies with the
ARGO-YBJ experiment.

\section{The "quasi-unaccompanied" events}

The ARGO-YBJ detector is constituted by a single layer of RPCs.
This carpet has a modular structure, the basic module being a
Cluster (7.6$\times$5.7 m$^2$), divided into 12 RPCs
(1.25$\times$2.80 m$^2$ each). Each chamber is read by 80 strips
of 61.8$\times$6.75 cm$^2$, logically organized in 10 independent
pads of 61.8$\times$55.6 cm$^2$ \cite{nim_argo}. The central
carpet, constituted by 10$\times$13 clusters with $\sim$93$\%$ of
active area, is enclosed by a guard ring partially instrumented
($\sim$40$\%$) in order to improve the identification of external
events. The full detector is composed by 154 clusters for a total
active surface of $\sim$6700 m$^2$. Due to the small pixel size
(the space information comes from strips and the time from the
pads) the detector is able to image the shower profile with an
unprecedented granularity.

The detector is not able to discriminate between different charged
particles and to identify charged hadrons. Anyway, the detector
granularity and the large continuous area allows to reconstruct in
great details small showers produced deep in the atmosphere, a few
g/cm$^2$ above the carpet. The idea is to select surviving primary
protons that interact only into the last interaction length above
the detector, thus producing small showers ("quasi-unaccompanied"
events) with a very steep lateral distribution. In fact, a very
collimated hadronic jet is produced and the electromagnetic
cascade originated by $\pi^0$'s is at very early stage of
development.

\begin{figure}
\begin{center}
\includegraphics [width=0.48\textwidth]{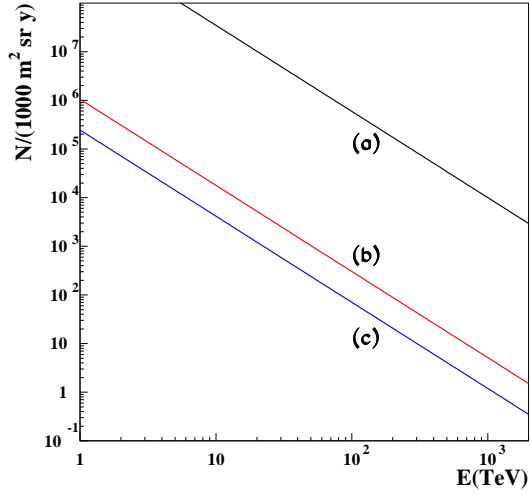}
\end{center}
\caption{The integral spectrum of surviving primary protons at the
YBJ atmospheric depth (c), of protons interacting into the last
interaction length (b) and of primary proton flux (a), for 1000
m$^2\cdot$year$\cdot$sr.}\label{fig-pexp}
\end{figure}

The surviving protons arriving at the YangBaJing atmospheric depth
suffer an attenuation given by N$_S$ = N$_0$ $\cdot$
e$^{-606/\lambda(E))}$, where $\lambda(E)$ is the nuclear
interaction mean free path. The number of expected events per 1000
m$^2\cdot$year$\cdot$sr in the energy range 1 TeV - 1 PeV is shown
in Fig.\ref{fig-pexp} compared with the primary protons flux
\cite{jacee}. The flux of protons interacting in the last
interaction length above the detector is also reported for
comparison. As it can be seen from the figure, about $3\cdot 10^4$
(500) events/year are expected above 10 (100) TeV.

In order to investigate the phenomenology of small EAS a number of
vertical proton-induced showers has been simulated by the
Corsika/QGSJet code \cite{corsika} with the first interaction
height fixed at the following value: 4350, 4400, 4500, 4600, 4800,
5000, 5500, 6000, 6500, 7000, 7500, 8000, 9000 m asl, for energies
ranging from 1 TeV to 1 PeV. We note that 4350 m asl corresponds
to 603 g/cm$^2$, 4400 m asl to 599 g/cm$^2$, 4500 m asl to 591
g/cm$^2$ and 4800 m asl corresponds to 568 g/cm$^2$. The shower
core positions have been uniformly sampled on a 100$\times$100
m$^2$ area centred on the detector.

The mean shower sizes for showers induced by vertical protons of
different energies are reported in Table \ref{tab1} as a function
of the first interaction height. From these values we can conclude
that a first interaction height higher than about 4500 m asl (591
g/cm$^2$) produces showers too big, indistinguishable from the
bulk of extensive air showers.
%
\begin{table}
\begin{center}
\begin{tabular}{cccccc}
\hline
E & 4350 & 4400 & 4500 & 4800 & 5500  \\
\hline
 10 TeV      & 41  & 57  & 94  & 257  & 1011 \\
 10$^2$ TeV  & 76  & 111 & 203 & 691  & 3632 \\
 10$^3$ TeV  & 139 & 225 & 464 & 1919 & 13786 \\
\hline
\end{tabular}
\caption{Mean shower size N$_{ch}$ (=
e$^{\pm}$+$\mu^{\pm}$+h$^{\pm}$) as a function of the proton
energy for different first interaction heights (m asl).
\label{tab1}}
\end{center}
\end{table}

\begin{figure*}
\vspace{-0.5cm}
\begin{minipage}[t]{.45\linewidth}
\includegraphics*[width=0.9\textwidth,angle=0,clip]{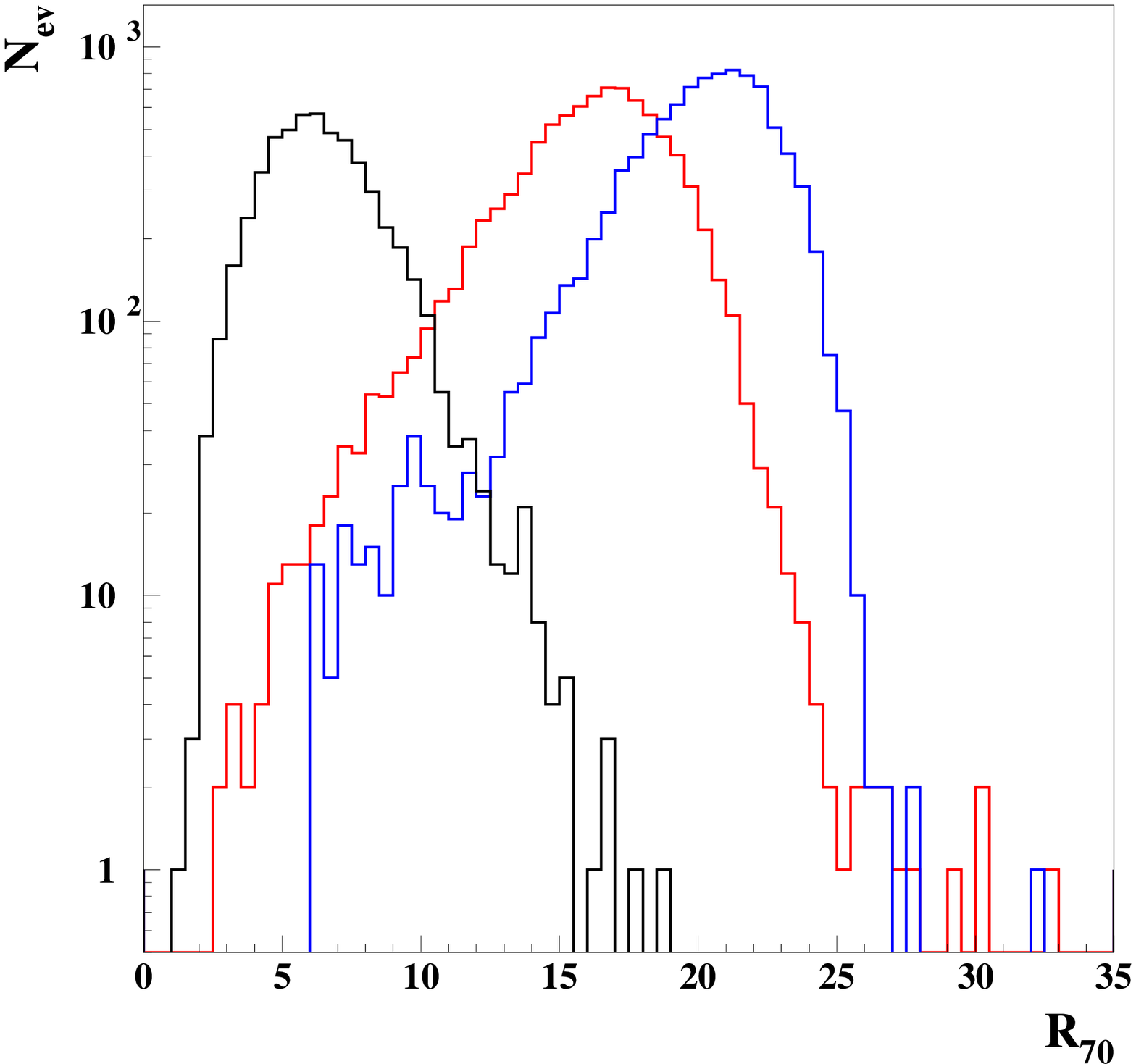}
\caption{\label{fig-r70} Distribution of the R$_{70}$ parameter
for different first interaction heights: 4350 (black), 4800 (red),
5500 (blue) m asl. The proton energy is 100 TeV.}
 \end{minipage}\hfill
\begin{minipage}[t]{.45\linewidth}
\includegraphics*[width=0.9\textwidth,angle=0,clip]{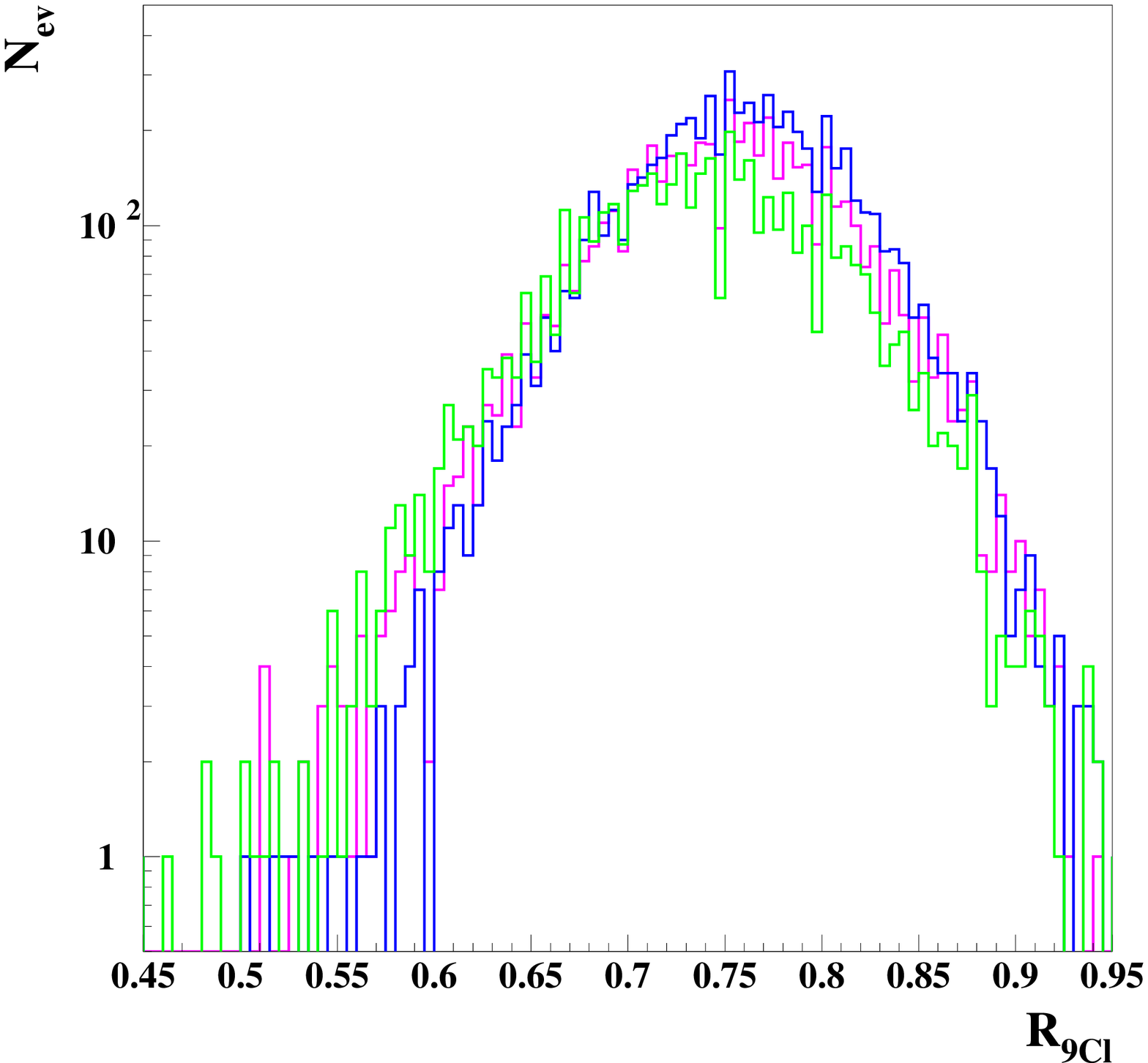}
\caption{\label{fig-n9cl} Distributions of the fraction R$_{9Cl}$
of fired strips inside A$_{9Cl}$ for different energies of the
primary proton: 10 (green), 100 (red), 1000 (blue) TeV. The first
interaction height is 4350 m asl. }
 \end{minipage}\hfill
\end{figure*}
%

\begin{figure*}
\vspace{-0.5cm}
\begin{minipage}[t]{.48\linewidth}
\includegraphics*[width=0.9\textwidth,angle=0,clip]{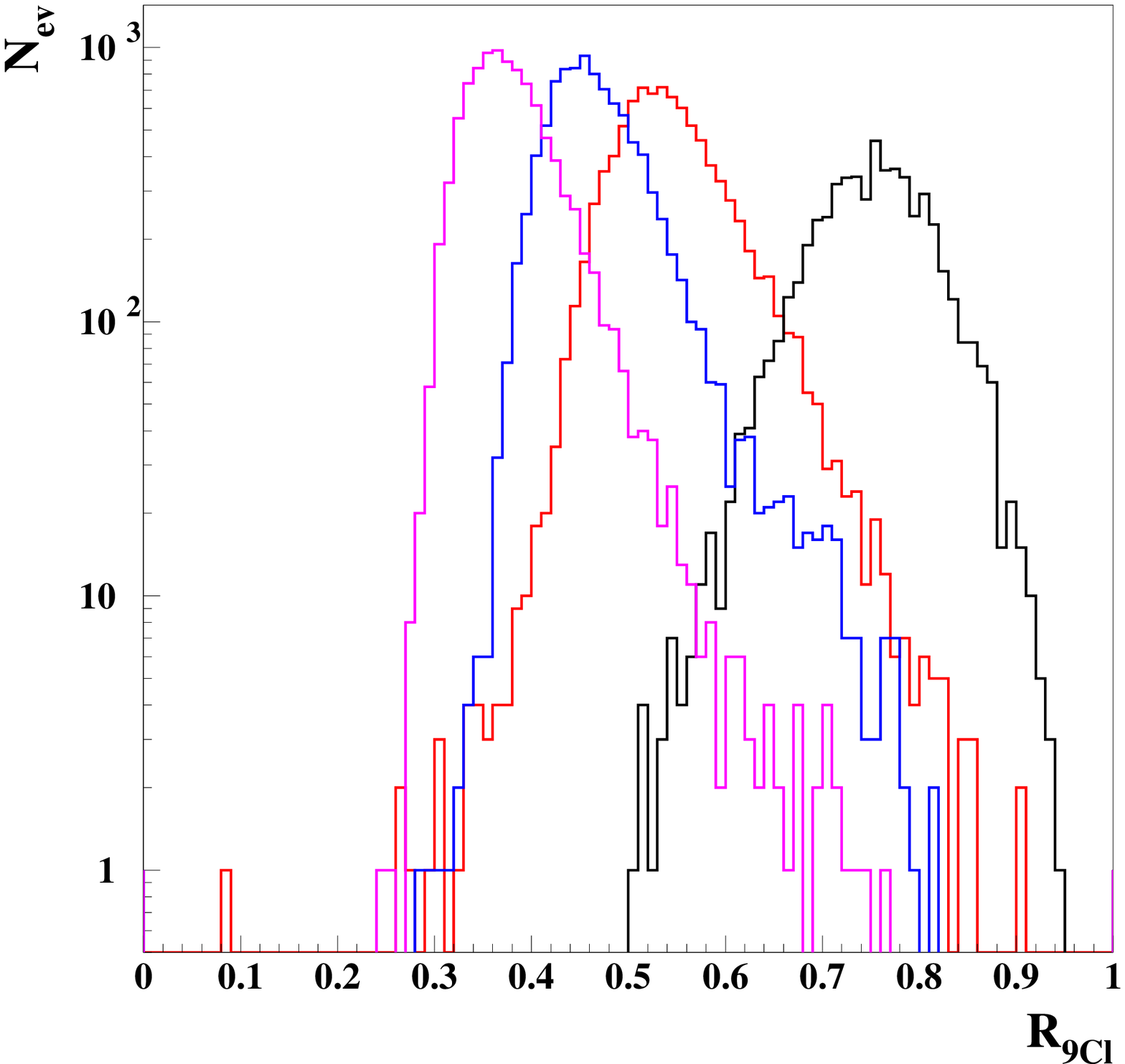}
\caption{\label{fig-n9cl1} Distributions of R$_{9Cl}$ for
different heights of the first interaction point: 4350 (black),
4800 (red), 5500 (blue), 6500 (cyan) m asl. The proton energy is
100 TeV.}
 \end{minipage}\hfill
\begin{minipage}[t]{.48\linewidth}
\includegraphics*[width=0.9\textwidth,angle=0,clip]{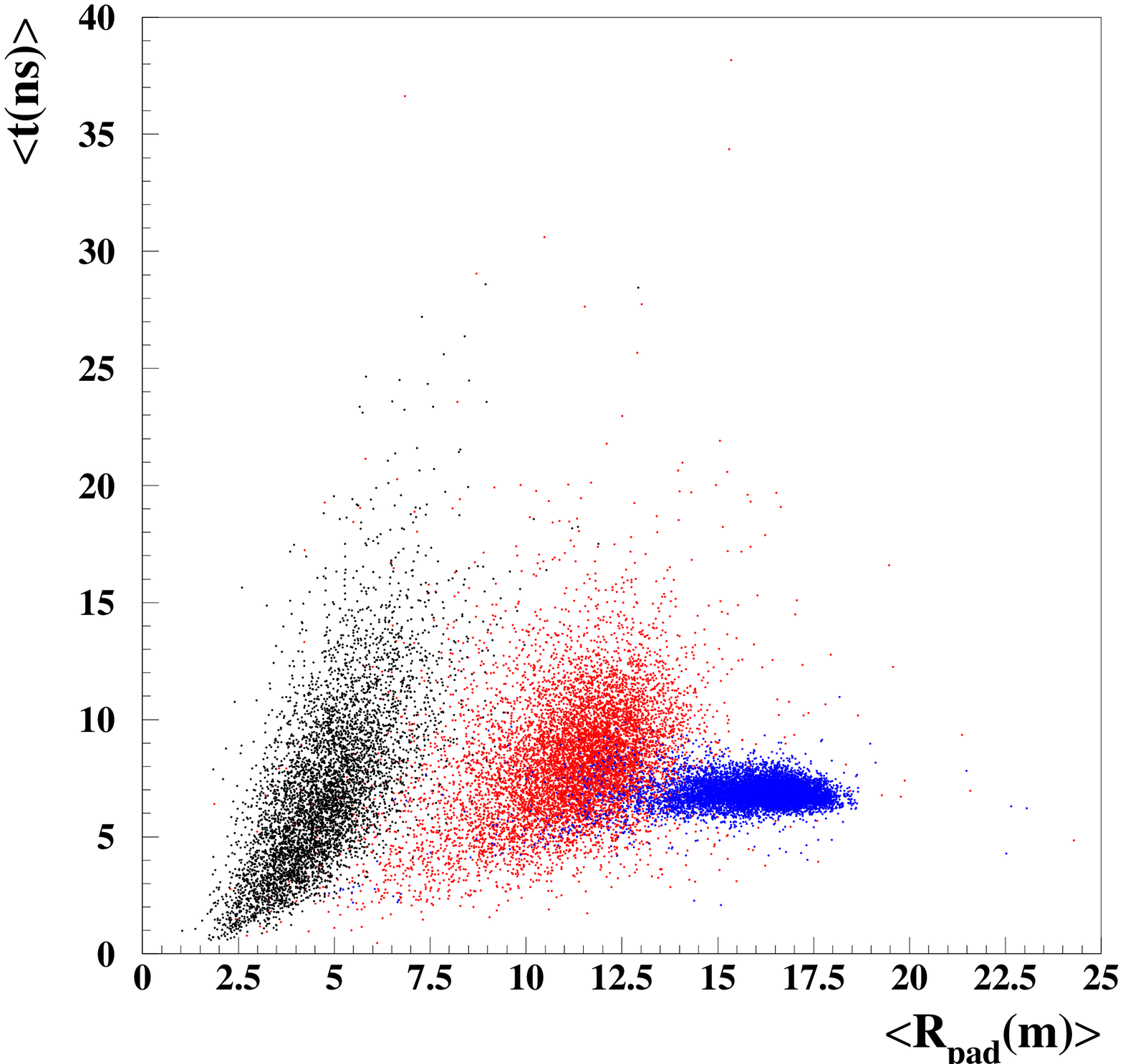}
\caption{\label{fig-time} Mean particle delays vs mean fired pad
distances from the shower core for 100 TeV protons interacting
4350 (black), 4800 (red), 5500 ( blue) m asl.}
 \end{minipage}\hfill
\end{figure*}

In order to investigate the "compactness" of the events, the
distributions of R$_{70}$, i.e., the radius around the core that
contains the 70$\%$ of the charged particles, is shown in
Fig.\ref{fig-r70} for 100 TeV proton-induced showers as a function
of the first interaction height. As expected, the events produced
a few g/cm$^2$ above the observation level have a lateral
extension very reduced being R$_{70}<$10-15 m. These values show
that the radial extension of such a showers is well contained in a
matrix constituted by 3$\times$3 Clusters (A$_{9Cl}$ =
22.8$\times$17.1 m$^2\sim$390 m$^2$) around the Cluster which
contains the highest strip multiplicity, suggesting to consider
the remaining carpet as an anti-coincidence area. In
Fig.\ref{fig-n9cl} the ratio R$_{9Cl}$=N$_{9Cl}$/N$_s$ between the
number N$_{9Cl}$ of fired strips inside the 3$\times$3 Cluster
matrix and the total number N$_s$ of fired strips is displayed for
different energies of protons interacting 4350 m asl. From the
figure it results that R$_{9Cl}$ is independent on the primary
energy in the range 10 - 1000 TeV.
%
In Fig.\ref{fig-n9cl1} the distributions of R$_{9Cl}$ for
different first interaction heights are shown. As expected, events
locally produced consist in very collimated jets with the charged
particles totally contained into a few meters around the shower
core position. By selecting showers for which R$_{9Cl}>$0.7 we
exclude the contribution of protons interacting above 450 g/cm$^2$
($\sim$6500 m asl), i.e. more than about 2 mean free paths above
the detector.

%

\section{The selection procedure}

The selection of protons interacting only a few g/cm$^2$ above the
observational level can be performed as follows: (1) selection of
events with the cluster with highest particle multiplicity inside
the inner 8$\times$11 clusters (A$_{f}\sim$3800 m$^2$) of the
central carpet, i.e. we exclude the outer ring constituted by 42
clusters; (2) selection of showers with R$_{9Cl}>0.7$; (3) the
shower core position of the selected events are reconstructed by
means of a weighted center of gravity method applied only inside
A$_{9Cl}$; (4) any core lying outside the fiducial area A$_f$ is
further rejected. The shower core position resolution of selected
events is less than 2 m. In this analysis only events with
N$_s>$30 have been taken into account.
%
\begin{table}
\begin{center}
\begin{tabular}{cccc}
\hline
f.i.h. (m asl) & 10 TeV & 10$^2$ TeV & 10$^3$ TeV \\
\hline
 4350 & 72$\%$ & 80$\%$ & 84$\%$  \\
 4400 & 34$\%$ & 41$\%$ & 42$\%$  \\
 4500 & 12$\%$ & 12$\%$ & 11$\%$  \\
 4600 &  6$\%$ &  6$\%$ &  4$\%$  \\
 4800 &  3$\%$ &  2$\%$ &  2$\%$  \\
 5000 &  2$\%$ &  1$\%$ & 0.5$\%$  \\
 5500 &  1$\%$ &  1$\%$ & 0.1$\%$  \\
\hline
\end{tabular}
\caption{Percentage of selected events a function of the proton
energy for different first interaction heights (f.i.h.). The
values are normalized to the showers with N$_{s}>$30.
\label{tab2}}
\end{center}
\end{table}
%
From the Table \ref{tab2} it results that the selection efficiency
is rather independent from the energy. With the described
procedure only a few percent of protons interacting above 4500 m
asl are selected. In Fig.\ref{fig-time} the mean particle delays
are plotted as a function of the mean fired pad distances from the
shower core (a measure of the shower lateral extension) for 100
TeV protons selected with the described procedure. As can be seen,
the degree of temporal profile curvature is higher for showers in
a earlier stage of development and can be used as a further
criterium for the proton selection.

\section{Conclusions}
The selection of surviving primary protons is possible with the
ARGO-YBJ experiment by measuring events interacting a few g/cm$^2$
above the observational level. The granularity of the detector
allows to investigate in great details the lateral and temporal
features of hadronic jets just produced.  Calculations are in
progress to evaluate the contribution of heavier primaries and to
properly take into account the fluctuations in the shower
development. A preliminary data analysis will be presented at the
conference.

\end{document}